# Genomic data processing with GenomeFlow

Junseok Park[1,2], Eduardo A. Maury[3,4], Changhoon Oh[5], Donghoon Shin[6], Danielle Denisko[1,7], and Eunjung Alice Lee[1,2]

[1] Division of Genetics and Genomics, Boston Children's Hospital, 3 Blackfan Circle, Boston, MA, USA
[2] Department of Pediatrics, Harvard Medical School, 25 Shattuck St, Boston, MA, USA
[3] Bioinformatics Integrative Genomics Program and Harvard/MIT MD-PHD Program, Harvard Medical School, Boston, MA, USA
[4] Program in Medical and Population Genetics, Broad Institute of MIT and Harvard, Cambridge, MA, USA
[5] Graduate School of Information, Yonsei University, 50 Yonsei-ro Seodaemun-gu, Seoul, Republic of Korea
[6] Department of Human Centered Design and Engineering, University of Washington, Seattle, WA, USA
[7] Department of Biomedical Informatics, Harvard Medical School, Boston, MA, USA

* ealice.lee@childrens.harvard.edu

## Abstract
Advances in genome sequencing technologies generate massive amounts of sequence data that are increasingly analyzed and shared through public repositories. On-demand infrastructure services on cloud computing platforms enable the processing of such large-scale genomic sequence data in distributed processing environments with a significant reduction in analysis time. However, parallel processing on cloud computing platforms presents many challenges to researchers, even skillful bioinformaticians. In particular, it is difficult to design a computing architecture optimized to reduce the cost of computing and disk storage as genomic data analysis pipelines often employ many heterogeneous tools with different resource requirements. To address these issues, we developed GenomeFlow, a tool for automated development of computing architecture and resource optimization on Google Cloud Platform, which allows users to process a large number of samples at minimal cost. We outline multiple use cases of GenomeFlow demonstrating its utility to significantly reduce computing time and cost associated with analyzing genomic and transcriptomic data from hundreds to tens of thousands of samples from several consortia. Here, we describe a step-by-step protocol on how to use GenomeFlow for a common genomic data processing task. We introduce this example protocol geared toward a bioinformatician with little experience in cloud computing.

## Background

During the last decade, large amounts of next-generation sequencing (NGS) data have been generated across a range of health and disease conditions [1]. For example, large consortia, such as The Cancer Genome Atlas (TCGA), International Cancer Genome Consortium (ICGC), Genotype-Tissue Expression (GTEx), Trans-Omics for Precision Medicine (TOPMED), and UK Biobank (UKBB) have created genomic, transcriptomic, epigenomic, and other types of omics data from hundreds of thousands of samples at an unprecedented resolution [2–5]. Due to their sheer volume, these data have been shared with the scientific community through cloud computing platforms. Pharmaceutical companies are heavily invested in mining these large-scale genomic data to gain insights into disease prevention, detection, and treatment [6]. While such therapy-driven investigations have been successful, analyzing available data at scale remains largely limited by the lack of off-the-shelf, multipurpose compute infrastructure for bioinformatic analyses he computing cost to set up and optimize the analyzes pipelines on large datasets poses a great burden on individual investigators and companies alike given the large amount of trial and error required.

To ameliorate this burden and to assist basic biology researchers in deploying algorithms and computational workflows on cloud platforms, a handful of methods have been developed. Tibanna is a tool that uses Amazon Web Service (AWS) as a cloud service provider, allowing users to define and run a scalable portable pipeline based on Common Workflow Language (CWL) or Workflow Description Language (WDL) [7]. Terra, by contrast, is a cloud-native platform that provides a genomic data analysis interface via user-defined WDL workflows on Google Cloud Platform (GCP) [8, 9]. Despite their utility and availability, these technologies still require multiple rounds of trial and error to select the most cost-effective computational resource parameters, which can quickly become too costly, especially when trying to deploy a novel algorithm (Table 1).

|  | Cloud Platform | Category | Development Language | Workflow Language | Cost Estimation | Cost Optimization |
| --- | --- | --- | --- | --- | --- | --- |
| GenomeFlow | GCP | Tool | Python | Snakemake, WDL | Yes | Yes |
| Terra.bio | GCP | Platform | - | WDL, CWL | No | No |
| Tibanna | AWS | Tool | Python | Snakemake, WDL, CWL | Yes | No |

**Table 1. Comparison between GenomeFlow, Terra.bio, and Tibanna.** Terra.bio is a platform native to the GCP cloud, while GenomeFlow and Tibanna are tools for GCP and AWS. Each platform and tool not only has unique features but also differs in functionality. WDL: Workflow Description Language, CWL: Common Workflow Language.

We therefore developed GenomeFlow, a tool that enables the deployment of resource-optimized scalable genomic pipelines on GCP [10]. Briefly, GenomeFlow implements scheduled task processing with failover functionality and scalable cloud computing architecture. Notably, It implements a resource optimizer to reduce cloud billing on massively parallel jobs[10].

Here, we describe two use cases to provide users with a reference pipeline. We additionally present a detailed GenomeFlow sample processing example for users without expert knowledge of cloud computing.

## Implementation

The main protocol of GenomeFlow is composed of two parts. The first is the distributional processing architecture design [11]. GenomeFlow distributes samples to user-defined pipelines using cloud computing resources. The second is the user-defined workflow. The workflow contains a pipeline that defines how to analyze input data by combining necessary tools. For this, the user provides a Dockerfile that contains an operational environment for the tools [12]. The Dockerfile is fed as input to GenomeFlow alongside a sample list. The sample list file should contain sample IDs. The user should also pass in the location where the samples will be downloaded as well as the tool to use for download. The controller of GenomeFlow passes a sample ID to each task in a job. The commands executed in each task must be defined by the user in the workflow. Users can define a single step or multiple steps for a task. Tasks with multiple steps can use different computing resources to optimize resource usage. To obtain the optimization parameters, the user can run the resource optimizer of GenomeFlow. In the case of downloading reference files, a user needs to upload their files to a Cloud Bucket and define the location of the workflow file. Finally, users must define the output location of the Cloud Bucket. GenomeFlow will store the result of each sample in the defined result location.

The workflow of GenomeFlow is configured according to a user-defined command and parameter set, so it is not limited to any specific kind of analysis. Any pipeline should work as long as the user-generated Dockerfile and command set are properly configured (Figure 1). GenomeFlow supports Python and uses WDL/Snakemake syntax for its workflow design. User-defined pipelines executed by GenomeFlow run on a Kubernetes cluster using an Airflow inspired controller and consists of a task scheduler to manage the status of jobs and individual tasks with a web-based log check interface on a MySQL database [11, 13, 14]. GenomeFlow acts as a tool to execute existing bioinformatics pipelines or novel tools in parallel, with an emphasis on optimizing cost configurations. We applied GenomeFlow to multiple genomic studies in our laboratory and were optimized able to optimize GCP computing resource usage, resulting in significantly reduced billing. In this paper, we show two use cases for a comprehensive overview of GenomeFlow, and we present a general scenario of its use.

**Figure 1. General protocol of GenomeFlow.** Users need to provide a workflow file using WDL or Snakemake syntax. Then, they can load the GenomeFlow library from Python and run it. (A) 1. Users need to create a Dockerfile, after which GenomeFlow will create a Docker image and place the image into the Google Cloud Platform (GCP) registry to deploy it in Google Kubernetes Environment (GKE) 2. Users need to upload the reference data to the bucket and set its location in the workflow file. 3. Sample ID file preparation 4. All the commands in the user-defined pipeline should be set in the workflow. (B) 1. GenomeFlow creates a bastion to build a Docker image, then uploads the image to the GCP container registry. 2. GenomeFlow is able to take in reference files locally if this setting is enabled by a user. 3. GenomeFlow inputs the sample IDs to the queue management service in GKE 4. GenomeFlow deploys the controller and prepares the controller's database for management purposes.

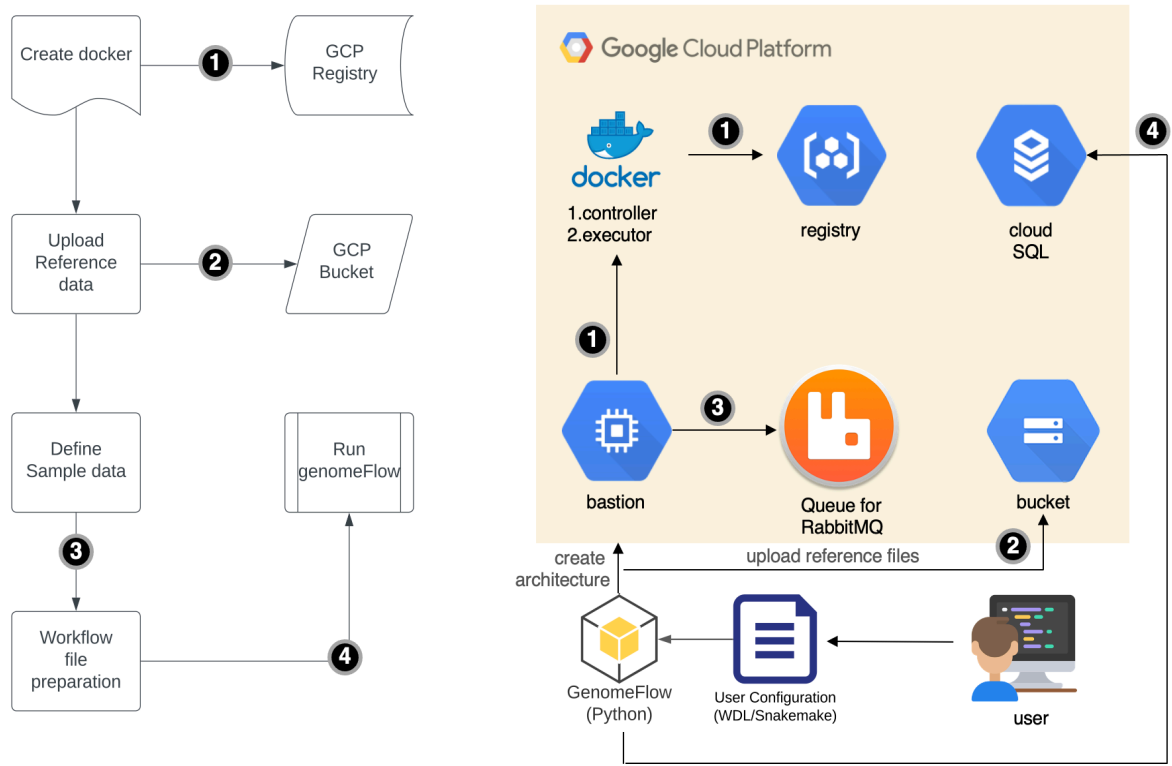

## Results

### Case report 1: Pan-cancer RNA-seq analysis to detect transposon-fusion RNA using rTea

We performed a pan-cancer analysis of 15,477 RNA-seq profiles from various human cancer and normal tissue samples using rTea, a computational method we developed to detect chimeric transcripts with transposon sequences [15]. We prepared three different sample groups using: (1) SRAtoolkit for GTEx (2) gdctool for TCGA and ICGC (3) and an on-premise downloader for CoPM [16, 17]. We uploaded necessary hg38 reference files to the GCP repository see [18]. We built a Dockerfile containing the rTea pipeline, a custom pipeline written in R (r-base:3.6.2) which executes various tools including fastp, hisat2, samtools, scallop, bamtools and bwa[19–24]. We prepared a one-step task for three jobs. Then, GenomeFlow injected the queue consumption module into the Dockerfile and ran all the tasks. Initially, we used 16 cores and 128 GB memory per instance without any queue management architecture. After using the recommended resource allocation parameters from GenomeFlow, we were able to reduce total memory from 128 GB to 64 GB, CPU from 16 cores to 8 cores, and storage from 500 GB to 250 GB. With GenomeFlow, we were able to use 4-8 cores and 16-32 GB memory per instance and reduced the disk storage requirement. The total running time was around 2 days for all samples including fail-over tasks. As a result, the average processing cost per sample using the n2-standard-4 configuration was $1.72, with an average processing time of around 7 hours per sample. This represents a 77% cost reduction compared to the original setup, which used the n2-highmem-16 configuration at a cost of $7.34 per sample (this cost may vary based on factors such as the GCP region and calculation date). Using GenomeFlow, we identified 52,227 cancer-specific fusions from the processed samples.

## Case report 2: Epilepsy panel-seq analysis to detect somatic mutations using MosaicHunter

We analyzed deep panel sequencing of 38 genes of interest known to be associated with epilepsy to identify somatic mutations using a revised version of MosaicHunter, originally developed for whole exome sequencing data [25]. Using GenomeFlow, we analyzed data from 30,789 samples from the Epi25k epilepsy cohort [26]. Similar to the rTea case, we prepared a one-step pipeline and ran all samples. We utilized GenomeFlow for MosaicHunter from the initial resource setup, thereby employing an optimized resource configuration of e2-standard-4 with 16 GB of memory and a 200 GB balanced persistent disk. The cost was approximately $0.12 per sample, with an average processing time of around 0.64 hours per sample. The cost may vary depending on GCP billing over time.

## Step-by-step use of GenomeFlow

We provide a step-by-step protocol of using GenomeFlow to process raw FASTQ files into analysis-ready BAM files drawing inspiration from Willet et al. [27] to allow users to easily utilize GenomeFlow for their own analytics projects on GCP. Users can also check the example Jupyter notebook on the GenomeFlow GitHub repository which is described in the code availability section.

### Equipment

Currently, GenomeFlow is only compatible with GCP. Although GCP has various resource types for their virtual machines, GenomeFlow only supports the following types of resources due to development limitations: e2-series, n2-series, n1-series [28]. A user is able to select a resource type for the machine; the default type is e2-standard-16 (16 vCPU and 64 GB Memory).

### Required software and hardware

Users need to install the gcloud CLI and get authorization with their Google ID before using GenomeFlow to access various GCP services. Users also need to install Docker Engine to build a Docker image. GenomeFlow uses Kubernetes to deploy tasks of a job defined by the workflow. The current supported version of Google Kubernetes Engine (GKE) is 1.23.8-gke.1900 [11]. GenomeFlow first creates the default nodes on the prepared GKE, then creates a service, queue (RabbitMQ), and controller (Flask), before finally deploying them to the default node of GKE [29, 30]. Afterwards, GenomeFlow creates nodes upon the user's request, and a pipeline in the workflow starts tasks as deployed pods in GKE nodes.

### Required data

Users need to download their GCP credentials as a JSON file from the GCP console and they need to provide a billing ID for their project creation on GCP. The user-defined files are as follows: reference file(s) and a sample ID list, a Dockerfile, and a workflow file.

### Reference file and sample ID file preparation

If a user's pipeline requires reference files, the user should consider uploading their files to a GCP bucket. The default size limit of the Docker container is 10 GB, which is not enough for large reference files. Users can indicate the GCP bucket location in the workflow file to skip the reference upload step, or users can use GenomeFlow to upload their files to a random GCP bucket location within the created project space. Stored reference files will automatically be transferred to the persistence disk (PD) of each pod (task), and tasks will use the reference files. In addition to the reference files, it is necessary to prepare a sample ID list file and sample file download commands including required download tools. These tools must be included in the user-built Dockerfile. The queue controller of GenomeFlow consumes the sample ID file to distribute sample IDs to each task.

**Dockerfile preparation**
GenomeFlow is composed of a GCP architecture creation function that runs a user-defined pipeline from a workflow file. Users must provide a Dockerfile since the workflow file uses a Docker image to run a pipeline. A user should write their own Dockerfile with compatible syntax to the specified version of Docker, and provide the Dockerfile location to the workflow file. The Dockerfile will be uploaded to a GCP bucket by GenomeFlow, and GenomeFlow will create a Docker image and transfer this image to a GCP container registry from the bastion instance. This Docker image is then deployed as a service pod in a GKE cluster (Docker version 20.10 of Ubuntu 20.04 LTS is installed in the bastion) from the bastion instance. Simultaneously, the address of the registry will be stored in the GenomeFlow controller database for the next step.

**Workflow file preparation**
The workflow file is a pipeline description file that prepares the GKE cluster on GCP and deploys tasks to a cluster node as specified. GenomeFlow supports two types of syntax: (1) Workflow Description language (WDL) and (2) Snakemake style language [31, 32]. In this example, we adopted Snakemake rules. The essential rules for GenomeFlow are as follows:

```
rule            : ruleparams
workdir         : stringliteral
configfile      : stringliteral
input           : parameter_list
output          : parameter_list
params          : parameter_list (including sampleID)
resources       : parameter_list
shell           : stringliteral
script          : stringliteral
metawrapper     : stringliteral
config          : stringliteral
image           : stringliteral
referencefile   : stringliteral
testsamplesize  : integer
```

The workflow file designates the steps to run a single task. A job consists of running the task across multiple samples, therefore corresponding to multiple tasks. Each task of GenomeFlow uses the concept of Directed Acyclic Graph (DAG) to represent its steps [13]. The tasks execute multiple steps and each step is assigned to a pod in a node of GKE. Each step can have different resource types based on the node type. This serves to optimize the resources per step and reduce costs. If a user does not allocate resources, GenomeFlow uses default resource types and downloads the default number of samples to find optimized resource parameters. Then, the user can check the optimized parameters and set them to reduce the cost.

**Executing the workflow**

GenomeFlow is a Python library. After preparing all of the required files, users can load GenomeFlow to execute the workflow. A procedure of GenomeFlow includes (1) loading the workflow file, (2) building the GCP architecture, (3) executing a test run to obtain resource optimization parameters, (4) running the pipeline, and (5) getting the results and removing the GCP environment. Below, we present a detailed outline of the workflow execution.

After installing GenomeFlow using pip, users load GenomeFlow:

```
>>> import genomeflow as gf
```

As an example, we will run a Fastq-to-BAM [27] pipeline. Steps include: sample download, read alignment and merging, and refinement of the BAM file.

Next, users load the workflow file along with other essential files, including the sample ID list and Dockerfile:

```
>>> workflow=gf.loadJobFile("exampleWorkflow.snakemake")
```

Users then create a GCP architecture for the pipeline in the workflow:

```
>>> gf.createArchitecture(workflow)
```

Optimized parameters (optParams) are returned from the test run of GenomeFlow. The optParams will replace the previous resource settings of the workflow.

```
>>> optParams=gf.findOptimizedParam(workflow)
```

Then, users run all the samples with optimized resource parameters. Users can check the progress through the GenomeFlow monitoring interface which preserves all generated logs from the GenomeFlow controller.

```
>>> gf.runPipeline(optParams)
```

**Get the results and check the costs**

Results are stored in a user designated GCP bucket location. Users are able to download each result per sample and check the file metadata using the GCP web interface. Users are also able to download all the data using gsutil of Cloud SDK with the following command:

```
$ gsutil cp -n -r [Bucket Address] [Destination Address]
```

After downloading all the sample results, users can remove GCP buckets and GCP architectures. The GCP architectures include GKE, Cloud SQL and GCP buckets, but do not include the GCP project itself.

```
>>> gf.removeProject(workflow)
```

If a user runs the remove function of GenomeFlow without any input parameters, it leaves the GCP project untouched. The purpose of preventing removal of the GCP project is to provide a cost check for the user. If users wish to remove the GCP project too, they can use the following command:

```
>>> gf.removeProject(workflow, all=True)
```

## Discussion

We developed GenomeFlow to optimize resource usage and reduce costs for bioinformaticians conducting parallel sample processing in a cloud computing environment. GenomeFlow generates a cloud architecture on GCP to process samples through user-defined Snakemake or WDL workflows using optimized cloud computing resource parameters, including memory, CPU, and storage configurations.

Despite the advancement that GenomeFlow provides for large-scale sample processing, users should be aware of its current limitations. In particular, GenomeFlow lacks automatic resource optimization for users. While GenomeFlow generates optimized parameters, users are still required to manually review and input these parameters to run their entire workflows, based on their specific requirements. This manual intervention adds an extra step to the workflow that may introduce potential errors or inefficiencies.

Another limitation of GenomeFlow is its platform dependency. Currently, GenomeFlow is designed to operate exclusively on GCP. Its automatic architecture preparation module relies on GCP APIs, restricting its compatibility with other cloud platforms. While the workflow language of GenomeFlow can theoretically be generalized to different cloud platforms, the automatic architecture preparation module is not interoperable, overall limiting its usability for users who operate on non-GCP environments.

To address these limitations, future developments of GenomeFlow will focus on enhancing user-friendliness and platform compatibility. Implementing features for automatic resource optimization, such as adaptive parameter tuning via machine learning, would streamline the workflow and reduce user burden. Additionally, expanding GenomeFlow's compatibility to support other cloud platforms, such as Amazon Web Services (AWS) or Microsoft Azure, would broaden its accessibility and utility for a wider range of users within the bioinformatics community [33, 34].

## Conclusion

In this work, we provide a step-by-step protocol for running GenomeFlow, a tool we developed to optimize cloud computing resource usage. Importantly, by optimizing resource usage, GenomeFlow greatly reduces costs, thus alleviating the huge burden on the bioinformatics community associated with processing hundreds to tens of thousands of samples. We show how users can easily obtain optimized resource parameters for running their custom pipelines on GCP and can monitor their jobs through a user-friendly web interface. We highlight two case reports in which GenomeFlow was employed to process RNA-seq and whole exome sequencing data from tens of thousands of samples, substantially reducing resource usage. While GenomeFlow was designed for GCP, future efforts will aim to expand its compatibility with other cloud platforms. Overall, we anticipate that GenomeFlow will be adopted widely to assist with a broad range of large-scale analyses across genomics and beyond.

## Availability and requirements
- Project name: GenomeFlow
- Project home page: https://github.com/ealeelab/genomeflow
- Operating system: Does not depend on operating system
- Programming language and version: Python 3.6 and 3.7
- License: GNU GPL v2, GNU GPL v3, CC BY-NC and MIT
- Any restriction to use by non-academics: No

## Declarations
**Ethics approval and consent to participate**
Not applicable

**Consent for publication**
Not applicable

**Availability of data and materials**
All the related code and Jupyter notebooks are available at https://github.com/ealeelab/genomeflow. Users can refer to the project's README.MD file for additional commands and instructions.

**Competing Interests**
The authors declare that they have no competing interests

**Fundings**
This work was supported by the NHGRI AnVIL Project (AnVIL AC3 Award 2022) and the National Research Foundation of Korea (NRF) funded by the Ministry of Education (2021R1A6A3A14046389).

**Authors' contributions**
JP developed GenomeFlow and analyzed data to demonstrate its performance. EM advised on the statistical analysis method. CO and DS developed the web interface of GenomeFlow and collected data for visualization. JP, DD, and EAL wrote the original manuscript, while EAL supervised the study. All authors participated in reviewing the final manuscript.

**Acknowledgment**
We thank Rich Fellmann, Ashrut Vora, and Andrew Carey of the Google Cambridge office for their help with technological issues on cloud computing platforms.